\documentclass[aps,prl,twocolumn,showpacs,superscriptaddress,nofootinbib]{revtex4}

\usepackage{graphicx}
\usepackage{amsmath}

\newcommand{\be}{\begin{eqnarray}}
\newcommand{\ee}{\end{eqnarray}}
\newcommand{\bea}{\begin{eqnarray}}
\newcommand{\eea}{\end{eqnarray}}

\newcommand{\ket}[1]{\mbox{$| #1 \rangle$}}

\newcommand{\proj}[1]{\mbox{$|#1\rangle \!\langle #1 |$}}

\def\H{{\cal H}}

\def\E{{\cal E}}

\def\d{{\mathrm d}}

\newcommand{\tr}{\mathop{\mathrm{tr}}\nolimits}
\renewcommand{\Im}{\mathop{\mathrm{Im}}}
\renewcommand{\equiv}{:=}
\def\circvert{\setbox0=\hbox{${\scriptscriptstyle\circ}{|}$}
              \vcenter{\hbox{${\scriptscriptstyle\circ}$}}\kern-.5\wd0 |}
\newcommand{\elabs}[1]{\circvert #1 \circvert}

\begin{document}

%%%%%%%%%%%%%%%%%%%%%%%%%%%%%%%%%%%%%%%%%%%%%%%%%%%%%%%%%%%%%%%%%%%%%%%%%%%%%%%

\title{Asymptotic entanglement capacity of the \\
       Ising and anisotropic Heisenberg interactions}

\author{Andrew M. Childs}
\affiliation{Center for Theoretical Physics,
             Massachusetts Institute of Technology,
             Cambridge, MA 02139, USA}
\affiliation{IBM T. J. Watson Research Center, P.O. Box 218,
             Yorktown Heights, NY 10598, USA}

\author{Debbie W. Leung}
\affiliation{IBM T. J. Watson Research Center, P.O. Box 218,
             Yorktown Heights, NY 10598, USA}

\author{Frank Verstraete}
\affiliation{SISTA/ESAT, Department of
Electrical Engineering, University of Leuven, Belgium}

\author{Guifr\'e Vidal}
\affiliation{Institute for Quantum Information, California Institute of
             Technology, Pasadena, CA 91125, USA}

\date[]{14 October 2002}

\begin{abstract}
We calculate the asymptotic entanglement capacity of the Ising interaction
$\sigma_z\otimes\sigma_z$, the anisotropic Heisenberg interaction
$\sigma_x\otimes\sigma_x + \sigma_y\otimes\sigma_y$, and more generally,
any two-qubit Hamiltonian with canonical form $K \!  = \! \mu_x \;
\sigma_x\otimes \sigma_x + \mu_y \; \sigma_y \otimes \sigma_y$.  We also
describe an entanglement assisted classical communication protocol using
the Hamiltonian $K$ with rate equal to the asymptotic entanglement
capacity.
\end{abstract}


\maketitle

%%%%%%%%%%%%%%%%%%%%%%%%%%%%%%%%%%%%%%%%%%%%%%%%%%%%%%%%%%%%%%%%%%%%%%%%%%%%%%%

The fundamental resource for quantum information processing is an
interaction between two quantum systems.  Any Hamiltonian $H_{AB} \neq H_A
+ H_B$ that is not a sum of local terms couples the systems $A$ and $B$.
Together with local operations, the coupling can be used to generate
entanglement \cite{DVCLP, BHLS, entanglement}, to transmit classical and
quantum information \cite{BHLS,BGNP,HVC,BS02}, and more generally, to
simulate the bipartite dynamics of some other Hamiltonian $H_{AB}'$ and
thus to perform arbitrary unitary gates on the composite space
$\H_{AB}=\H_A\otimes \H_B$ \cite{simulationplusgates,BCLLLPV,cata}.

Much experimental effort has been devoted to creating entangled states of
quantum systems, including those in quantum optics, nuclear magnetic
resonance, and condensed matter physics \cite{For}.  Determining the
ability of a system to create entangled states provides a benchmark of the
``quantumness'' of the system.  Furthermore, such states could ultimately
be put to practical use in various quantum information processing tasks,
such as superdense coding~\cite{Bennett92} or quantum
teleportation~\cite{Bennett93}.

The theory of optimal entanglement generation can be approached in
different ways.  For example, Ref.~\cite{DVCLP} considers {\em
single-shot} capacities.  In the case of two-qubit interactions, and
assuming that ancillary systems are not available, Ref.~\cite{DVCLP}
presents a closed form expression for the entanglement capacity and
optimal protocols by which it can be achieved.  In contrast,
Ref.~\cite{BHLS} considers the {\em asymptotic} entanglement capacity,
allowing the use of ancillary systems, and shows that when ancillas are
allowed, the single-shot and asymptotic capacities are in fact the same.
However, such capacities can be difficult to calculate because the
ancillary systems may be arbitrarily large.

In this paper, we calculate the asymptotic entanglement capacity of any
two-qubit interaction that is locally equivalent to $\mu_x \;
\sigma_x\otimes \sigma_x + \mu_y \; \sigma_y \otimes \sigma_y$, and thus
present a connection between the results of Refs.~\cite{DVCLP} and
\cite{BHLS}.  We consider the use of ancillary systems, and show that they
do not increase the entanglement capacity of these interactions.  Thus in
these cases, the asymptotic capacity discussed in Ref.~\cite{BHLS} is in
fact given by the expression presented in Ref.~\cite{DVCLP}.  As an
application of this result, we present an explicit ensemble for
entanglement assisted classical communication \cite{BHLS}, implicitly
found in Ref.~\cite{BS02}, at a rate equal to the entanglement capacity.
We also give an alternative ensemble achieving the same rate.  Finally, we
conclude by presenting some numerical results on the entanglement capacity
of general two-qubit interactions.

We begin by reviewing some definitions and known results.  Let
$\ket{\psi}$ be a state of the systems $A$ and $B$.  This state can always
be written using the Schmidt decomposition \cite{Peres}, 
\be
  \ket{\psi} \equiv \sum_i \sqrt{\lambda_i} ~ 
                           \ket{\phi_i}_A\otimes\ket{\eta_i}_B
\,,
\ee 
where $\{\ket{\phi_i}\}$ and $\{\ket{\eta_i}\}$ are orthonormal sets of
states, and $\lambda_i > 0$ with $\sum_i \lambda_i = 1$.  The entanglement
between $A$ and $B$ is defined as
\be
  E(\ket{\psi}) \equiv - \sum_i \lambda_i\log \lambda_i 
\,.
\ee 
(Throughout this paper, the base of $\log$ is $2$.)

Reference \cite{DVCLP} considers maximizing the rate of increase of
entanglement when a pure state is acted on by $e^{-iHt}$, the evolution
according to a time-independent Hamiltonian $H$ (we set $\hbar=1$
throughout this paper).  We refer to this maximal rate as the {\em
single-shot} entanglement capacity.  When no ancillas are used, this is
given by 
\be
  E_{H}^{(1*)} \equiv \max_{\ket{\psi} \in \H_{AB}} 
  \lim_{t\rightarrow 0}\frac{E( e^{-i H t} \ket{\psi}) -E(\ket{\psi})}{t} 
\,.
\label{eq:EHstar}
\ee 
Here the rate of increasing entanglement is optimized over all possible
pure initial states of $\H_{AB}$ without ancillary systems.  In fact, the
single-shot capacity may be higher if ancillary systems $A'$ and $B'$, not
acted on by $H$, are used.  For this reason, we may consider the
alternative single-shot entanglement capacity
\be
  E_{H}^{(1)} \equiv \sup_{\ket{\psi} \in \H_{AA'BB'}} 
  \lim_{t\rightarrow 0}\frac{E( e^{-i H t} \ket{\psi}) -E(\ket{\psi})}{t}
\,.
\label{eq:EH}
\ee 
Note that in Eqs.~(\ref{eq:EHstar}) and (\ref{eq:EH}), the limit is the
same from both sides even though it might be the case that $E_H^{(1*)} \ne
E_{-H}^{(1*)}$ in general (and similarly for $E_H^{(1)}$).

For any two-qubit Hamiltonian $H$, Ref.~\cite{DVCLP} shows that it is
locally equivalent to a {\em canonical form}
\be
  \sum_{i=x,y,z} \mu_i \; \sigma_i \otimes \sigma_i \,, \quad
  \mu_x \geq \mu_y \geq |\mu_z|
\,.
\ee
In terms of this canonical form, the optimal single-shot entanglement
capacity of any two-qubit interaction without ancillas is given by
\bea
  E_H^{(1*)}  &=& \alpha (\mu_x + \mu_y) \,, \\
  \alpha &\equiv& 2\max_x \sqrt{x \, (1\!-\!x)} 
  \, \log \! \left(\frac{x}{1\!-\!x}\right)
          \approx 1.9123
\,,
\label{eq:noanc}
\eea
where the maximum is obtained at $x_0 \approx 0.9168$.  In addition,
$E_H^{(1)}$ may be strictly larger than $E_H^{(1*)}$ when $|\mu_z| > 0$
\cite{DVCLP}.

Reference \cite{BHLS} considers the {\em asymptotic} entanglement capacity
$E_H$ for an arbitrary Hamiltonian $H$. $E_H$ is defined as the maximum
average rate at which entanglement can be produced by using many
interacting pairs of systems, in parallel or sequentially.  These systems
may be acted on by arbitrary collective local operations (attaching or
discarding ancillary systems, unitary transformations, and measurements).
Furthermore, classical communication between $A$ and $B$ and possibly
mixed initial states are allowed.
Reference \cite{BHLS} proves that the asymptotic entanglement capacity in
this general setting turns out to be just the single-shot capacity in 
Ref.~\cite{DVCLP}, $E_H = E_{H}^{(1)}$, for all $H$, so
\be
  E_{H} = \sup_{\ket{\psi} \in \H_{AA'BB'}}
  \lim_{t\rightarrow 0}\frac{E( e^{-i H t} \ket{\psi}) -E(\ket{\psi})}{t}
\label{eq:EHA}
\,.
\ee 
Note that the definition of the capacity involves a supremum over both all
possible states and all possible interaction times, but in fact it can be
expressed as a supremum over states and a limit as $t \to 0$, with the
limit and the supremum taken in either order.

Let $\ket{\psi}$ be the optimal input in Eq.~(\ref{eq:EH}) or
(\ref{eq:EHA}).  When $\ket{\psi}$ is finite dimensional, the entanglement
capacity can be achieved \cite{DVCLP,BHLS} by first inefficiently
generating some EPR pairs, and repeating the following three steps: ($i$)
transform $nE(\ket{\psi})$ EPR pairs into $\ket{\psi}^{\otimes n}$
\cite{BPPS,dilute}, ($ii$) evolve each $\ket{\psi}$ according to $H$ for a
short time $\delta t$, and ($iii$) concentrate the entanglement into
$n(E(\ket{\psi})+\delta tE_{H})$ EPR pairs \cite{BPPS}.

In this paper, we show that $E_K^{(1)} = E_K^{(1*)}$ 
for any two-qubit Hamiltonian with canonical form 
\be
  K \equiv \mu_x \; \sigma_x\otimes \sigma_x 
         + \mu_y \; \sigma_y \otimes \sigma_y \,, \quad
  \mu_x \geq \mu_y \geq 0 
\,, 
\label{eq:xy}
\ee
so that all three entanglement capacities are equal:
\be
  E_K = E_K^{(1)} = E_K^{(1*)}
\,.
\ee
The optimal input is therefore a two-qubit state, and the optimal protocol 
applies.  
In particular, for these Hamiltonians, which include the Ising
interaction $\sigma_z \otimes \sigma_z$ and the anisotropic Heisenberg
interaction $\sigma_x\otimes\sigma_x + \sigma_y\otimes\sigma_y$,
entanglement can be optimally generated from a $2$-qubit initial state
$\ket{\psi}$ without ancillary systems $A'B'$.  As mentioned above,
this result is not generic, since ancillas increase the amount of
entanglement generated by some two-qubit interactions, such as the
isotropic Heisenberg interaction $\sigma_x\otimes\sigma_x +
\sigma_y\otimes\sigma_y + \sigma_z\otimes\sigma_z$ \cite{DVCLP}.

In the following, we will focus on computing the asymptotic entanglement
capacity of the interaction
\be
  K_{xx}\equiv \sigma_x\otimes\sigma_x
\,.
\label{eq:xx}
\ee
One way to see that this is sufficient to determine the asymptotic
entanglement capacity of $K$ in Eq.~(\ref{eq:xy}) is to note that $K$ is
{\em asymptotically equivalent} to
\be
  K' \equiv (\mu_x+\mu_y) \sigma_x\otimes\sigma_x
\label{eq:x}
\ee
and that $E_{t H} = |t| E_{H}$ for two-qubit Hamiltonians.  The asymptotic
equivalence of $K$ and $K'$ is based on the following two facts: ($i$)
$K'$ and fast local unitary transformations on qubits $A$ and $B$ can
simulate $K$ \cite{BCLLLPV}; conversely, ($ii$) the Hamiltonian $K$ can be
used to simulate $K'$ given a {\em catalytic} maximally entangled state,
without consuming the entanglement of $A'B'$, which subsequently can be
re-used \cite{cata}.  Therefore, Hamiltonians $K$ and $K'$ are
asymptotically equivalent resources given local operations and an
asymptotically vanishing amount of entanglement.  In particular, $E_K =
E_{K'}$.  This equivalence could be generalized to other capacities, but
for the specific case of entanglement capacity, a simpler proof is
available.  The simulation ($i$), which does not require a catalyst,
demonstrates $E_K \le E_{K'}$.  After computing $E_{K'}$, we will see that
the protocol of Ref.~\cite{DVCLP} saturates this bound, so in fact $E_K =
E_{K'}$ with no need for ancillas to achieve either capacity.

We now present the optimization of Eq.~(\ref{eq:EHA}) for $K_{xx}$.  We
suppose that in addition to the qubits $A$ and $B$ on which $K_{xx}$ acts,
$d$-dimensional ancillas $A'$ and $B'$ are used, where $d$ is arbitrary.
We can always write the Schmidt-decomposed initial state $\ket{\psi}$ as
\bea
  \ket{\psi}
    &=& \sum_i \sqrt{\lambda_i} \; 
		\ket{\phi_i}_{AA'} \otimes \ket{\eta_i}_{BB'}
        \label{eq:Schmidt} \\
    &=& (U\otimes V)(\sqrt\Lambda \otimes I_{BB'}) \ket{\Phi} \\
    &=& U \sqrt\Lambda V^T \otimes I_{BB'} \ket{\Phi} \label{eq:Schmidt2}
\,,
\eea
where $U$ and $V$ are unitary matrices on $\H_{AA'}$ and $\H_{BB'}$,
$\Lambda$ is a diagonal matrix with diagonal elements $\Lambda_{ii} =
\lambda_i$, $\ket{\Phi} = \sum_i \ket{i}_{AA'}\otimes \ket{i}_{BB'}$, and
we have used the fact that
\be
  (I \otimes M) \ket{\Phi} = (M^T \otimes I) \ket{\Phi}
\label{eq:trick}
\ee  
for any operator $M$.  Defining $\rho \equiv \tr_{BB'} \proj{\psi}$, the
entanglement capacity of any Hamiltonian $H$ is
\bea
  E_H
    &=& \sup_{\ket{\psi}} \tr\left( -\frac{\d\rho}{\d t} \log \rho 
                        - \rho \frac{\d\log \rho}{\d t} \right) \nonumber\\
    &=& \sup_{\ket{\psi}} \tr\left( -\frac{\d\rho}{\d t} \log \rho \right)
\,.
\eea
The variation of $\rho$ can be computed using perturbation theory
\cite{DVCLP}:
\be
  \!\!
  \frac{\d\rho}{\d t} = -i \tr_{BB'} [H,\proj{\psi}]
                      = 2 \Im \tr_{BB'}(H \proj{\psi})
\,.
\label{eq:pert}
\ee
Letting $R = U \sqrt\Lambda V^T$ and considering $H=K_{xx}$, we have
\bea
   && \tr_{BB'} \left( K_{xx}\proj{\psi} \right) \nonumber\\
  &=& \tr_{BB'} \left[(X \otimes X)(R \otimes I_{BB'})
      \proj{\Phi}(R^\dag \otimes I_{BB'}) \right] \nonumber \\
  &=& \tr_{BB'} \left[(XRX^T \otimes I_{BB'})
      \proj{\Phi}(R^\dag \otimes I_{BB'}) \right]  \nonumber \\
  &=& X R X^T R^\dag
\,,
\label{eq:ptrace}
\eea
where we have introduced $X \equiv \sigma_x \otimes I$, with the identity
operator acting on the ancilla.  The first equality follows simply from
substitution of $K_{xx}$ and $\ket{\psi}$ by their expressions in
Eqs.~(\ref{eq:xx}) and (\ref{eq:Schmidt2}); the second uses
Eq.~(\ref{eq:trick}); and the third employs the fact that for any
operators $M_1, M_2$, 
\be
  \tr_{BB'} [(M_1 \otimes I_{BB'}) \proj{\Phi} (M_2 \otimes I_{BB'})]
  = M_1 M_2
\,.
\ee
Since $\rho = U \Lambda U^\dag$, we have
\be
  E_{K_{xx}}
    = \sup_{\ket{\psi}} \tr\left( -U^\dag \frac{\d\rho}{\d t} U 
      \log \Lambda \right)
\,.
\label{eq:entk}
\ee
Using Eqs.~(\ref{eq:pert}) and (\ref{eq:ptrace}), and introducing the
Hermitian operators $X_U = U^\dag X U$ and $X_V = V^\dag X V$, we have
\be
  U^\dag \frac{\d\rho}{\d t} U = 2 \Im X_U \sqrt\Lambda X_V^T \sqrt\Lambda
\,.
\ee
Letting $U,V,\Lambda$ attain the supremum in Eq.~(\ref{eq:entk}) (up to an
amount that can be made arbitrarily small), we find
\bea
  E_{K_{xx}}
    &=& -2 \Im \tr \left( X_U \sqrt\Lambda X_V^T
        \sqrt\Lambda \log\Lambda \right) \nonumber\\
    &=& i \tr \left[ (X_U \sqrt\Lambda X_V^T - X_V^T \sqrt\Lambda X_U)
                       \sqrt\Lambda \log\Lambda \right] \nonumber\\
    &=& i \tr \left[ M (X_U \circ X_V) \right]
\,,
\label{eq:had}
\eea
where we have introduced the real, skew-symmetric matrix
\be
  M_{ij} \equiv \sqrt{\lambda_i \lambda_j} 
                \log (\lambda_j/\lambda_i)
\,,
\ee
and the symbol $\circ$ denotes the Hadamard (i.e., element-wise) product
of matrices.  In the second line of Eq.~(\ref{eq:had}) we have used
\be
  \Im \tr A = \tr (A-A^\dag) / 2i
\ee
and the fact that $\Lambda$, $X_U$, and $X_V$ are Hermitian.  The last
line can be checked by explicitly writing the trace in terms of matrix
elements.

From Eq.~(\ref{eq:had}) we obtain the following upper bound for
$E_{K_{xx}}$ (here $\elabs{A}$ denotes the element-wise absolute value,
i.e., $\elabs{A}_{ij} = |A_{ij}|$): 
\bea
  E_{K_{xx}}
    &\leq& \tr(\elabs{M} \; \elabs{X_U \circ X_V}) \nonumber\\
    &\leq& \sup_{P} \tr(\elabs{M} P) \nonumber\\
    &\leq& 2 \max_x\sqrt{x(1-x)}\log[x/(1-x)] \nonumber\\
       &=& \alpha \approx 1.9123
\,,
\label{eq:bounds}
\eea
where $P$ is a permutation operator and $x \in [0,1]$.  The first line
uses the triangle inequality.  The second inequality follows from noticing
that $\elabs{X_U \circ X_V}$ is a doubly substochastic matrix
\cite{Bhatia}.  Indeed, for any two complex numbers $v$ and $w$ one has
that $2|vw| \leq |v|^2 + |w|^2$, and consequently, for any two unitary
matrices $V$ and $W$, 
\bea
  \sum_i|V_{ij}W_{ij}|\leq\sum_i(|V_{ij}|^2+|W_{ij}|^2)/2=1,\nonumber\\
  \sum_j|V_{ij}W_{ij}|\leq\sum_j(|V_{ij}|^2+|W_{ij}|^2)/2=1, 
\eea
which implies that the matrix $\elabs{V\circ W}$, with entries
$|V_{ij}W_{ij}|$, is doubly substochastic. Therefore a doubly stochastic
matrix $Q$ exists such that  $|X_U \circ X_V|_{ij} \leq Q_{ij}$ for all
$i$ and $j$ \cite{Bhatia}, so that $\tr (\elabs{M} \; \elabs{X_U \circ
X_V}) \leq \tr (\elabs{M} Q)$. But $Q$ is a convex combination of
permutation operators $P_k$, $Q= \sum_k p_k P_k$, which implies that
$\tr\elabs{M} Q \leq \sup_{P} \tr(\elabs{M} P)$.  Finally, the third
inequality in Eq.~(\ref{eq:bounds}) follows from noticing that
\bea
  |M|_{ij} &=& \sqrt{\lambda_i \lambda_j}
               |\log(\lambda_j/\lambda_i)| \nonumber\\
    &=& (\lambda_i + \lambda_j)
        \sqrt{\frac{\lambda_i}{\lambda_i+\lambda_j}
              \frac{\lambda_j}{\lambda_i+\lambda_j}}
	|\log(\lambda_j/\lambda_i)| \nonumber \\
 &\leq& (\lambda_i + \lambda_j)
        \max_{x}\sqrt{x(1-x)}\log[x/(1-x)] \nonumber\\ 
    &=& (\lambda_i + \lambda_j) \alpha/2
\,,
\eea
and that
\be
  \tr(\elabs{M}P) \leq \frac{\alpha}{2} \sum_{ij} (\lambda_i+ \lambda_j)P_{ij} 
  = \alpha \sum_i \lambda_i = \alpha
\,,
\ee
where we have used the facts that $P$ is a permutation matrix and that
$\sum_i \lambda_i = 1$.  Comparison of Eqs.~(\ref{eq:noanc}) and
(\ref{eq:bounds}) shows that, indeed, $E_{K_{xx}}=E^{(1*)}_{K_{xx}}$,
completing the proof.

We have shown that ancillary systems are not needed when optimizing
entanglement generation by any two-qubit Hamiltonian with canonical form
given by Eq.~(\ref{eq:xy}). More specifically, there is a universal
optimal two-qubit initial state given by \cite{DVCLP}
\be
  \ket{\psi_{\max}} \equiv \sqrt{x_0} \ket{0}_A\otimes\ket{1}_B 
                      - i \sqrt{1-x_0} \ket{1}_A \otimes \ket{0}_B
\,.
\label{eq:psimax}
\ee

As an application of the above, we discuss how to use the Hamiltonian $K$
to enable classical communication between Alice and Bob.  This has been
studied in \cite{BHLS}, in which the entanglement assisted forward
classical capacity $C_{\rightarrow}^{E}$ (the maximum rate for the
Hamiltonian $H$ to communicate from Alice to Bob when free, unlimited
shared entanglement is available) is shown to be
\be
  \!\!\! C_{\rightarrow}^{E}(H) = \sup_{\E} \left[ \lim_{t\rightarrow 0} 
  \frac{\chi(\tr_{AA'}e^{-iHt}\E)-\chi(\tr_{AA'}\E)}{t} \right] \! ,
\label{eq:CE}
\ee
where $\E=\{p_i,\ket{\psi_i}\}$ is an ensemble of bipartite states,
$e^{-iHt}\E$ and $\tr_{AA'}\E$ denote the respective transformed ensembles
$\{p_i,e^{-iHt}\ket{\psi_i}\}$ and $\{p_i, \tr_{AA'}\proj{\psi_i}\}$, and 
\be
  \chi(\{p_i,\rho_i\}) \equiv 
  S\left(\sum_i p_i \rho_i\right) - \sum_i p_i S(\rho_i)
\ee
is the Holevo information of the ensemble $\{p_i,\rho_i\}$, where $S$ is
the von Neumann entropy.  Reference \cite{BHLS} also describes a protocol
to achieve the rate in the bracket of Eq.~(\ref{eq:CE}) for any ensemble
$\E$.  

For any two-qubit Hamiltonian $H$, Ref.~\cite{BS02} constructs an ensemble
with communication rate $E_H$, which implies $C_{\rightarrow}^{E}(H) \geq
E_H$.  This ensemble, which is not necessarily optimal, is defined in
terms of an optimal state for entanglement generation.  This ensemble
$\E_1$ can now be made more explicit for Hamiltonian $K$ in light of our
findings:  
\bea
 p_1&\equiv&\frac{1}{2},~ \ket{\psi_1} 
 \equiv \sqrt{x_0} \ket{0}_A\!\otimes\!\ket{1}_B 
 + i \sqrt{1\!-\!x_0} \ket{1}_A\! \otimes \!\ket{0}_B, \nonumber \\
 p_2&\equiv&\frac{1}{2},~ \ket{\psi_2} 
  \equiv \sqrt{x_0} \ket{0}_A\!\otimes\!\ket{0}_B 
  - i \sqrt{1\!-\!x_0} \ket{1}_A \!\otimes \!\ket{1}_B, \nonumber
\eea
where $x_0$ is defined after Eq.~(\ref{eq:noanc}). 
For ensemble $\E_1$ we find
\bea
  &&\chi(\tr_A \E_1) = S(I/2) - S(\tr_A\proj{\psi_1}) 
                     = 1 - E(\ket{\psi_{\max}})\nonumber \\
  &&\chi(\tr_A(e^{-i\delta t K}\E_1)) 
                     = 1 - [E(\ket{\psi_{\max}})-\delta t E_K]
\eea
and therefore the net rate at which classical bits are transmitted is
indeed $\Delta \chi/\delta t=E_K$.

Next we present an alternative ensemble $\E_2$ of product states with
the same communication rate:
\bea
  p_1&\equiv&\frac{1}{2},~ \ket{\psi_1} \equiv \frac{1}{\sqrt{2}}
    \left(\begin{array}{c} 1 \\ 1 \end{array}\right)_A\!\!\otimes
    \left(\begin{array}{c} \sqrt{x_0} \\ -i\sqrt{1-x_0} \end{array}\right)_B
  \nonumber \\
  p_2&\equiv&\frac{1}{2},~ \ket{\psi_2} \equiv \frac{1}{\sqrt{2}}
    \left(\begin{array}{c} 1 \\ -1 \end{array}\right)_A\!\!\otimes
    \left(\begin{array}{c} \sqrt{x_0} \\ i\sqrt{1-x_0} \end{array}\right)_B.
  \nonumber
\eea
Here, we use $K$ to simulate $K'$ \cite{cata}, under which the ensemble
evolves. For ensemble $\E_2$, $S(\tr_A\proj{\psi_i})=0$, so
\bea
  &&\chi(\tr_A \E_2) = H_2(x_0) \nonumber\\
  &&\chi(\tr_A(e^{-i\delta t K}\E_2)) = H_2\left(x_0-2 \delta t
                                        \sqrt{x_0(1-x_0)}\right) \nonumber\\
  &&\hspace*{17.9ex} = H_2(x_0) + E_K \delta t
\eea
(where $H_2$ is the binary entropy).  Thus the communication rate is again
$\Delta \chi/\delta t=E_K$.

The main difference between these two ensembles is that the states in
ensemble $\E_1$ are entangled whereas the states in ensemble $\E_2$ are
not.  In the first case the interaction $K$ is used to decrease the degree
of entanglement between Alice and Bob or, equivalently, to make the states
of Bob's ensemble $\tr_A \E_1$ less mixed and thus more distinguishable.
The same increase of distinguishability for the pure states of Bob's
ensemble $\tr_A\E_2$ is achieved by conditionally rotating them with $K$,
in a way that they become more orthogonal to each other.  We note, in
addition, that ensembles $\E_1$ and $\E_2$ can be prepared using different
remote state preparation techniques \cite{rsp}.

\begin{figure}
\includegraphics[width=3.3in]{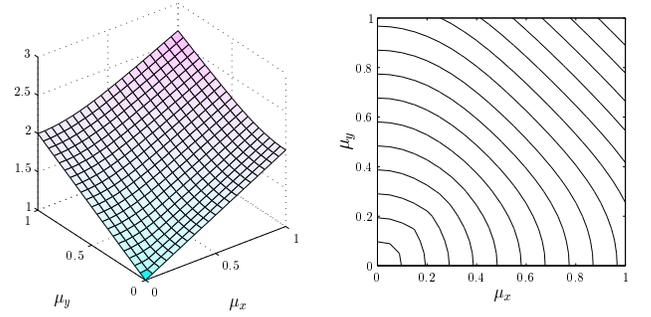}
\caption{Numerically optimized entanglement capacity of the two-qubit
Hamiltonian $\mu_x \; \sigma_x \otimes \sigma_x + \mu_y \; \sigma_y
\otimes \sigma_y + \sigma_z \otimes \sigma_z$ with single qubit ancillas
on each side.  The vertical axis in the left figure is in units of
$\alpha$.}
\label{fig:qubit}
\end{figure}

In conclusion, we have computed the asymptotic entanglement capacities of
all two-qubit Hamiltonians that are locally equivalent to $\mu_x \;
\sigma_x\otimes \sigma_x + \mu_y \; \sigma_y \otimes \sigma_y$ by showing
that this capacity can be achieved without the use of ancillas.  However,
as discussed above, ancillas are necessary to achieve the capacity in
general.  Although we do not have a closed form expression for the
capacity of an arbitrary two-qubit Hamiltonian, we can present partial
results in this direction.  The numerically optimized entanglement
capacity of a general two-qubit Hamiltonian is shown in
Fig.~\ref{fig:qubit}.  Numerically, we find that the optimum can be
achieved with single-qubit ancillas on both sides.  For Hamiltonians of
the form $K_{\mu_{xy}} = \mu_{xy}(\sigma_x\otimes \sigma_x + \sigma_y
\otimes \sigma_y) + \sigma_z \otimes \sigma_z$, we conjecture that the
entanglement capacity is given by
\be
  E_{K_{\mu_{xy}}} \! = 2 \max \big\{
    &\!\!\!\!\!\!\!&
     \sqrt{p_1 p_2} \log(p_1/p_2) \, [\sin\theta + \mu_{xy} \sin(\varphi-\xi)] 
     \nonumber\\
   +&\!\!\!\!\!\!\!&
     \sqrt{p_2 p_4} \log(p_2/p_4) \, [\sin\varphi + \mu_{xy} \sin(\theta-\xi)] 
     \nonumber\\
   +&\!\!\!\!\!\!\!&
      \sqrt{p_1 p_4} \log(p_1/p_4) \, \mu_{xy} \sin \xi
  \big\}
\ee
where the maximum is taken over $p_1>0$, $p_2>0$, $p_4=1-p_1-2p_2>0$, and
$\theta,\varphi,\xi \in [0,2\pi)$.  This expression was found by
investigating the structure of the numerical optimum, and it agrees well
with the numerical results.  It does not seem possible to simplify this
expression further, which suggests that in general, capacities may not
have simple closed form expressions, but can only be expressed as
maximizations of multivariable transcendental functions.  Nevertheless, it
would be useful to show that this maximization can be taken over a finite
number of parameters by proving an upper bound on the dimension of the
ancillas.

%%%%%%%%%%%%%%%%%%%%%%%%%%%%%%%%%%%%%%%%%%%%%%%%%%%%%%%%%%%%%%%%%%%%%%%%%%%%%%%

\medskip

We thank Aram Harrow, Patrick Hayden, and John Smolin for interesting
discussions.  We also thank Michael Nielsen for comments on the
manuscript.
AMC received support from the Fannie and John Hertz Foundation.  DWL was
supported in part by the NSA under ARO Grant No.\ DAAG55-98-C-0041.  FV
thanks John Preskill and the Caltech IQI for their hospitality.  GV is
supported by the US National Science Foundation under Grant No.\
EIA-0086038.
This work was supported in part by the Cambridge--MIT Foundation, by the
Department of Energy under cooperative research agreement
DE-FC02-94ER40818, and by the National Security Agency and Advanced
Research and Development Activity under Army Research Office contract
DAAD19-01-1-0656.

%%%%%%%%%%%%%%%%%%%%%%%%%%%%%%%%%%%%%%%%%%%%%%%%%%%%%%%%%%%%%%%%%%%%%%%%%%%%%%%

\end{document}